\documentclass[aps, prl, reprint, superscriptaddress]{revtex4-1}

\usepackage{graphicx}
\usepackage{mathrsfs,amsmath,bm}
\usepackage{cleveref}
\usepackage{xcolor}
 \DeclareMathOperator{\Real}{Re} % declare matrix trace operator

\begin{document}
\title{Microscopic theory of in-plane critical field in two-dimensional Ising superconducting systems}
\author{Hongchao Liu}
\affiliation{International Center for Quantum Materials, Peking University, Beijing 100871, China}
\author{Haiwen Liu}
\email{haiwen.liu@bnu.edu.cn}
\affiliation{Center for Advanced Quantum Studies, Department of Physics, Beijing Normal University, Beijing 100875, China}
\author{Ding Zhang}
\affiliation{State Key Laboratory of Low-Dimensional Quantum Physics, Department of Physics, Tsinghua University, Beijing 100084, China}
\affiliation{Beijing Academy of Quantum Information Sciences, Beijing 100193, China}
\author{X. C. Xie}
\affiliation{International Center for Quantum Materials, Peking University, Beijing 100871, China}
\affiliation{Beijing Academy of Quantum Information Sciences, Beijing 100193, China}
\affiliation{CAS Center for Excellence in Topological Quantum Computation, University of Chinese Academy of Sciences, Beijing 100190, China}
\date{\today}

\begin{abstract}
We study the in-plane critical magnetic field of two-dimensional Ising superconducting systems, and propose the microscopic theory for these systems with or without inversion symmetry. Protected by certain specific spin-orbit interaction which polarizes the electron spin to the out-of-plane direction, the in-plane critical fields largely surpass the Pauli limit and show remarkable upturn in the zero temperature limit. The impurity scattering and Rashba spin-orbit coupling, treated on equal-footing in the microscopic framework, both weaken the critical field but in qualitatively different manners. The microscopic theory is consistent with recent experimental results in stanene and Pb superconducting ultra-thin films.
\end{abstract}
\date{\today}
%%% \pacs{71.10.-w, 71.15.Rf, 71.55.Ak}
\maketitle

\textit{Introduction.---} The pair breaking mechanisms of a conventional superconductor, such as scattering with paramagnetic impurities\cite{AG1961} as well as generation of vortices\cite{PhysRev.147.295}, have been intensively studied\cite{fulde50years}. In layered superconductors, the reduction of dimensionality weakens the orbital effect when the magnetic field parallels the layered plane\cite{tinkham1996}, therefore providing a possible route to a large in-plane critical field $B_c$. However, due to the Zeeman energy splitting, the Cooper pairs in conventional superconductors normally become unstable when the magnetic field exceeds the Pauli limit\cite{chandrasekhar1962,PhysRevLett.9.266}.
In contrast, the translational symmetry breaking Fulde-Ferrell-Larkin-Ovchinnikov (FFLO) state can stabilize Cooper pairs beyond the Pauli limit\cite{FF,LO}, with
 the requirement that the superconductor locating in the clean limit\cite{Shimahara2007,FFLOReview}. Moreover, the spin-orbit scattering (SOS) randomizes the spin orientation by weakening the paramagnetism effect, as shown in the Klemm-Luther-Beasley (KLB) theory, and leads to enhancement of $B_c$\cite{PhysRevB.12.877}. Recent studies on two-dimensional (2D) crystalline superconductors\cite{IwasaNR,Iwasa2015,Wang2015,lu2015,saito2016,xi2016,PhysRevX.8.021002,YePNAS,HuntNC,Wang2018} have pointed out yet a third mechanism to enhance $B_c$,  originating from the spin-orbit coupling of the system. The in-plane inversion symmetry breaking leads to out-of-plane polarization of electron spin, and the Cooper pairing is protected against the in-plane magnetic field\cite{lu2015,saito2016,xi2016}.
 
 The impurity scattering may randomize the spin orientation, and thus renormalize the in-plane critical field $B_c$. Moreover, apart from the aforementioned inversion-asymmetric Ising superconducting systems, certain inversion-symmetric 2D materials can host spin splitting around the $\Gamma$ point due to intrinsic spin-orbit coupling\cite{PhysRevLett.111.136804,liao2018,Xu2019}. New microscopic model is needed to investigate the pairing breaking mechanism in these inversion-symmetric systems. And the combination effects of spin-orbit coupling and impurity scattering need to be studied on equal footing. Such investigations, so far have not been proposed, can give quantitative explanations for the enhancement of in-plane $B_c$ in Ising superconductors, including the recent discoveries of inversion-symmetric Ising superconductiviting systems\cite{falson2019type,Wang2019}.

%Both of the aforementioned mechanisms predict enhanced $B_c$ that should be over 1000 T due to the large spin splitting, which has never been seen experimentally. A more realistic model therefore has to take into account the impurity scattering and the Rashba type SOI---both may suppress the enhancement of $B_c$, and need to be.
% Such a quantitative analysis is still lacking.
% On the other hand, the impurity scattering randomize the spin direction and the Rashba-type SOI polarizes spin to the in-plane direction, which both suppress the Ising pairing and need to be properly considered in a quantitative manner.
  	
In this paper, we provide general microscopic analysis for 2D Ising superconductors, and treat the Ising-inducing intrinsic SOI, the impurity scattering, and the Rashba SOI simultaneously.
Starting from a schematic physical analysis, we propose a microscopic model and derive the in-plane critical field relation $B_c(T)$ for inversion-symmetric Ising superconductivity and inversion-asymmetric one respectively.
The comparison of theoretical results with recent experiments is also given, with a remarkable upturn at the ultra-low temperature regime, which is qualitatively different from the KLB formula.

\begin{figure}
		\includegraphics[width=0.48\textwidth]{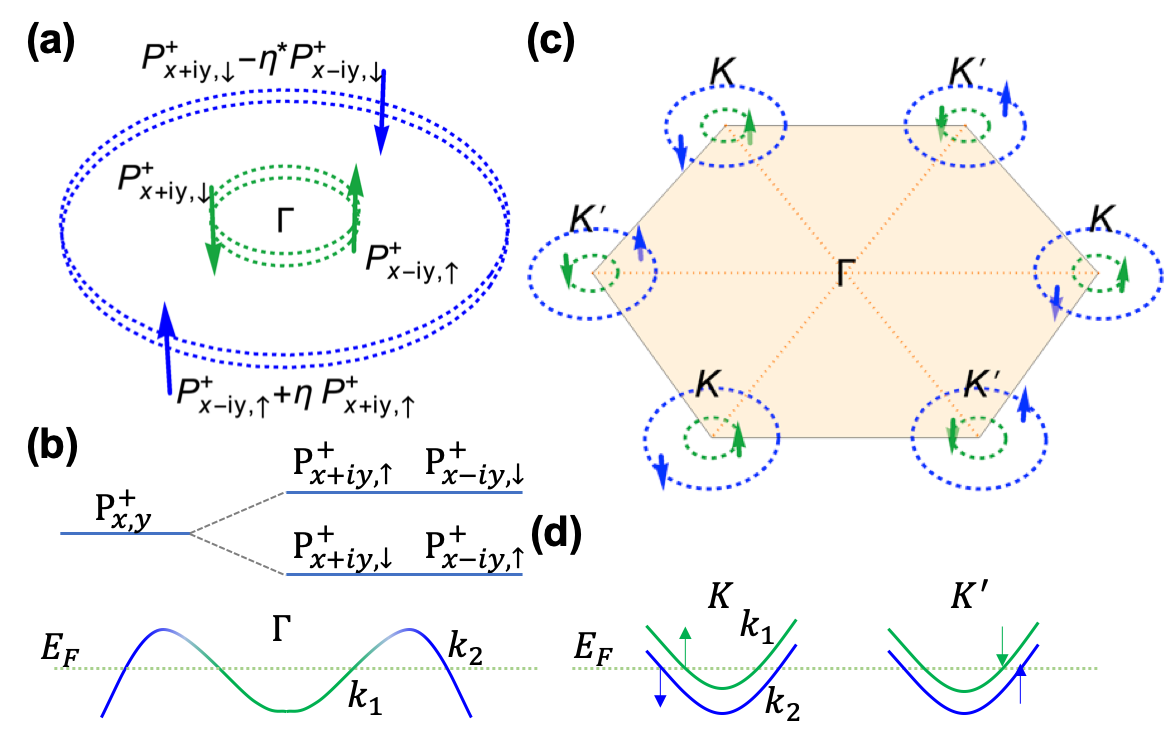}
		\caption{(Color online) Schematic diagrams of inversion-asymmetric and inversion-symmetric Ising superconductivity. (a) Two doubly-degenerate FSs (green and blue dashed circle-pairs) around the $\Gamma$ point in the Brillouin zone of stanene (inversion-symmetric Ising). The arrows on the FSs denote spin directions along $\pm z$-direction, $k_1$ in green and $k_2$ in blue. Each FS is doubly degenerate, consisting of two different $P_{x,y}^+$ orbitals and out-of-plane spin directions. The pairing happens between $P_{x-iy,\uparrow}^+$ and $P_{x+iy,\downarrow}^+$ on the same FS if $k_r$ is small ($\eta$ can't be neglected at large $k_r$ and is explained in \textit{Discussions}).
		(b) 4-fold degenerate $P_{x,y}^+$ level of stanene splits to two doubly-degenerate levels due to SOI. The lower level forms two FSs at $k_1$ (green) and $k_2$ (blue).
		(c) Brillouin zone of ultrathin Pb film (inversion-asymmetric Ising), with double non-degenerate FSs at each valley (green and blue dashed circles). Intervalley Cooper pairing form between electrons with the same color and opposite momentum.
		(d) Valley structure of Pb film in the vicinity of $E_{F}$. Each valley has two electron pockets, and Zeeman-type SOI polarizes electron spin oppositely around $K$ and $K'$.}
		\label{Fig schematic}
	\end{figure}

\textit{Inversion-symmetric Ising superconductivity.---} This type of 2D superconductivity happens in a system with its energy valley at zero-momentum point $\Gamma$ and two doubly-degenerate Fermi surfaces (FSs) around it, where the SOI is not Zeeman-type.
For example, in few-layer stanene, intrinsic SOI splits the 4-fold degenerate $P_{x,y}^+$ level into two doubly-degenerate levels, opening a gap at the $\Gamma$ point\cite{PhysRevLett.111.136804,liao2018}.
The lower level crosses $E_{F}$ at two different Fermi wavevectors $k_1, k_2$, forming two different-shaped FSs\cite{liao2018}, each of which holds two states [see Fig. \ref{Fig schematic}(a) and (b)].
If $k_r$ is small, the two eigenstates of the higher energy level can be approximated by $P_{x+iy,\uparrow}^+, P_{x-iy,\downarrow}^+$, while those of the lower level can be approximated by $P_{x-iy,\uparrow}^+, P_{x+iy,\downarrow}^+$.
Therefore the SOI at valley $\Gamma$ can be viewed as an out-of-plane magnetic field which takes opposite value
$-B_{eff}\hat{z}$ and $B_{eff}\hat{z}$ on different orbits $P_{x+iy}^+$ and $P_{x-iy}^+$ respectively ($\hat{z}$ is a unit vector perpendicular to the plane).
In this way the system has time-reversal symmetry (TRS) at zero $B$, and electrons with opposite momenta and spins on the same FS can form Cooper pairs.
The s-wave pairing with orbit-locked out-of-plane spin can give rise to the large in-plane $B_c$.
	
	The double-FS structure of inversion-symmetric Ising case differs from those of inversion-asymmetric Ising in their shape and location. Based on previous study\cite{Xu2019}, the system can be represented by a four band model with basis $(P_{x+iy,\uparrow}^+, P_{x-iy,\uparrow}^+, P_{x-iy,\downarrow}^+, P_{x+iy,\downarrow}^+ )$ to describe the normal state stanene with external in-plane magnetic field $B\hat{x}$
		\begin{align}
			H_{II}(\bm{k})  & = A k^2 + \left[ \begin{array}{c c}
				H_+(\bm{k})   & \ -\mu_BB\sigma_x  \\ -\mu_BB\sigma_x  & \ H_- (\bm{k})
				\end{array} \right] , \label{stanene model} \\
			H_{\pm}(\bm{k}) & = \left[ \begin{array}{c c}
				M_0-M_1k^2   & \ \  v (\pm k_x-i k_y)  \\ v (\pm k_x+i k_y)  & \ \  -M_0+M_1k^2
				\end{array} \right] ,
		\end{align}
with $A, M_0, M_1, v $ as fitting parameters, and $\mu_B$ as the effective Bohr magneton. At $B=0$, $H_{II}$ has TRS and two dispersion relations
	\begin{equation}
		E_\pm(k)=Ak^2\pm \sqrt{( M_0-M_1k^2 )^2+v ^2k^2} ,
	\end{equation}
	each of which is doubly degenerate. At small $k$, the degenerate eigenstates of $E_-(k)$ can be approximated by $P_{x-iy,\uparrow}^+, P_{x+iy,\downarrow}^+$. We consider the lower band $E_-(k)$ crosses $E_{F}$ at two different Fermi wavevector $k_1$ and $k_2$ [see Fig. \ref{Fig schematic}(b)], and take into account the spin-independent scattering disorder within each FS by setting a mean free time $\tau_0$ in the Green's function\cite{abrikosov1962,PhysRev.147.295,supmatthis}. The critical field for each FS can be solved within the Werthamer-Helfand-Hohenberg (WHH) framework\cite{PhysRev.147.295,supmatthis}, and be joined together in light of quasiclassical two-band Usadel equations\cite{PhysRevB.67.184515,supmatthis}. The critical field satisfies:
		\begin{multline}\label{is-eq}
			\frac{2w}{\lambda_0} F(\tilde{m}_1,t,b) F(\tilde{m}_2,t,b) + \bigg(1+\frac{\lambda_-}{\lambda_0}\bigg) F(\tilde{m}_1,t,b) \\
			+ \bigg(1-\frac{\lambda_-}{\lambda_0}\bigg) F(\tilde{m}_2,t,b) =0 ,
		\end{multline}
where $r=1, 2$ labels the two bands, $\lambda_{rr'}$ is the matrix of BCS coupling constants (assumed $\lambda_{11}>\lambda_{22}$), $\lambda_{\pm} =\lambda_{11}\pm\lambda_{22},\ \lambda_0 =\sqrt{\lambda_-^2+4\lambda_{12}^2}, \ w = \lambda_{11}\lambda_{22}-\lambda_{12}^2$,
	\begin{multline} \label{is-eqF}
		F(\tilde{m}_r,t,b) \equiv \ln t + \frac{b^2}{\tilde{m}_r^2 +b^2} \\
		\times \bigg[ \Real \psi \bigg(\frac{1}{2}+\frac{i\sqrt{\tilde{m}_r^2 +b^2}}{2\pi t}\bigg) -\psi\bigg(\frac{1}{2}\bigg) \bigg] ,
	\end{multline}
	\begin{equation} \label{para2}
		\tilde{m}_r = \frac{ \sqrt{ (M_0-M_1k_r^2)^2 + v ^2k_r^2 }}{k_BT_{c}+\hbar/(2\pi \tau_0) },\ t= \frac{T}{T_{c} },\ b=\frac{ \mu_BB_c}{ k_BT_{c}},
	\end{equation}
$\psi(x)$ is digamma function, $\tilde{m}_r$ is effective SOI, and all the functions and parameters appearing in Eq. (\ref{is-eq}) and (\ref{is-eqF}) are dimensionless.
When the two FSs have analogous shapes ($\tilde{m}_1\approx\tilde{m}_2$), or interlayer coupling is very weak ($\lambda_{12}\ll \lambda_-$), Eq. (\ref{is-eq}) has a simpler one-band form $F(\tilde{m}_1,t,b)=0$.
Meanwhile, in more generic situation, influenced by both bands, the Eq. (\ref{is-eq}) can't be simplified, with the curve of inversion-symmetric Ising case deviating from $F(\tilde{m}_1,t,b)=0$.

\begin{figure}
		\includegraphics[width=8.6cm]{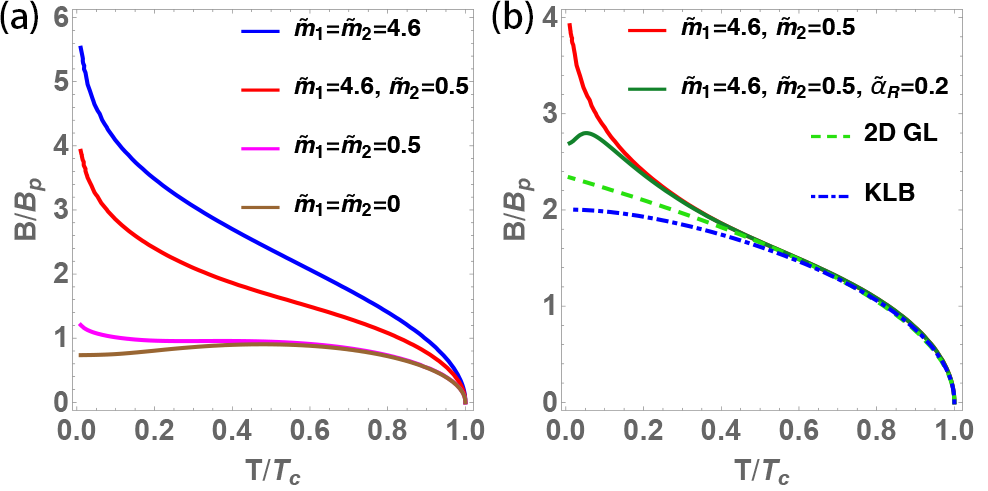}
		\caption{(Color online) In-plane critical field normalized by $B_{p}$ as a function of temperature normalized by $T_{c}$. (a) Inversion-symmetric Ising theory Eq. (\ref{is-eq}) (red curve) and its special cases $\tilde{m}_1=\tilde{m}_2\neq0$ (blue, pink), and zero SOI case $\tilde{m}_1=\tilde{m}_2=0$ (brown). The common parameters not showing in the legend are taken as $\lambda_{11}=3, \lambda_{22}=1, \lambda_{12}=2$. (b) Inversion-symmetric Ising (red), inversion-symmetric Ising with Rashba-type SOI (dark green), 2D GL theory, and KLB theory. $\tilde{\alpha}_R$ is the effective Rashba SOI, and parameters of 2D GL and KLB theory are $\xi_{GL} d_{sc}=\frac{0.88\hbar}{eB_{p}}, \tau_{so}=9.9 k_{B}T_{c}$. % Remark to Hongchao: the lineshape of 2D-GL and KLB need to be changed to dotted and dashed line for better print.
		}
		\label{is-fig}
	\end{figure}

    Near $T_{c}$, the inversion-symmetric Ising formalism is consistent with  the 2D Ginzburg-Landau (GL) theory\cite{PhysRev.129.2413}  and the KLB theory\cite{PhysRevB.12.877}. As shown in Fig. \ref{is-fig}(b) we compare the inversion-symmetric Ising case with 2D GL theory\cite{PhysRev.129.2413}
		\begin{equation}\label{2D GL}
			B_c =\frac{\sqrt{12}}{2\pi} \frac{\Phi_0 }{\xi_{GL} d_{sc}}\sqrt{1-\frac{T}{T_{c}}},
		\end{equation}
and KLB theory of SOS mechanism\cite{PhysRevB.12.877,PhysRevB.25.171}
		\begin{equation}\label{KLB}
			\ln \frac{T}{T_{c}} + \psi \bigg(\frac{1}{2}+\frac{3\tau_{so}}{2\hbar}  \frac{\mu_B^2 B_c^2}{2\pi k_BT} \bigg)-\psi\bigg(\frac{1}{2}\bigg)=0.
		\end{equation}
Here $\Phi_0, \xi_{GL}, d_{sc}, \tau_{so}$ denote flux quantum, GL coherence length at zero temperature, effective thickness of superconductivity, and SOS time, respectively. When the temperature is near $T_{c}$, $B_c$ predicted in Eqs. (\ref{is-eq})(\ref{KLB}) are both proportional to $\sqrt{1-T/T_{c}}$, consistent with 2D GL theory.
		
	In low temperature region, inversion-symmetric Ising theory has a remarkable upturn, and apparently larger field than 2D GL, KLB theory, and zero SOI case. This upturn establishes a stark difference from the standard pair-breaking (SPB) theory discussed by P. Fulde\cite{fulde1970,fulde1979,fulde50years}. Various TRS-breaking cases lead to the same equation and similar thermodynamics properties, and TRS-breaking factors function as a generic parameter in that equation. If TRS-breaking factor is the field $B_c$, and the equation of SPB theory is
	\begin{equation}\label{spb}
		\ln \frac{T}{T_{c}} + \Real \psi \bigg(\frac{1}{2}+\frac{\hbar}{\tau_\mathcal{R}(B_c)}\frac{1}{2\pi k_BT } \bigg)-\psi\bigg(\frac{1}{2}\bigg)=0,
	\end{equation}
where $\tau_\mathcal{R}(B_c)$ is a function of $B_c$ acting as the generic TRS-breaking parameter with a real or complex value. From Eq. (\ref{spb}), one can obtain $\lim\limits_{T\rightarrow 0}\left|\tau_\mathcal{R}(B_c)\right| = \frac{2e^\gamma}{\pi} \frac{\hbar}{k_BT_{c}}$, where $\gamma=0.577$ is the Euler constant. If we set $\left|\tau_\mathcal{R}(B_c)\right|=\frac{\hbar}{\mu_B B_c}$ or $\left|\tau_\mathcal{R}(B_c)\right|=\frac{2}{3\tau_{so}}(\frac{\hbar}{\mu_B B_c})^2$ in Eq. (\ref{spb}), we reach zero SOI case $F(0,t,b)=0$ or KLB theory Eq.(\ref{KLB}). Therefore the critical field near $T=0$ in those two situation is asymptotic to some finite constant and no upturn can happen. However, Eq. (\ref{is-eq}) shows remarkable upturn in low temperature region, which is indeed a distinguished experimental property of inversion-symmetric Ising superconductivity\cite{falson2019type}. Although the FFLO state also shows upturn $B_{c}$ in the low temperature regime, however the impurity scattering can destroy the possible FFLO states\cite{Shimahara2007,FFLOReview}. Moreover, the upturn feature in the low temperature regime is quantitatively different from the 2D FFLO state\cite{supmatthis}.
	
	The upturn can be explained from the two-level structure of the four-band model Eq. (\ref{stanene model}). In Fig. \ref{Fig schematic}(b), we plot the schematic diagram of energy levels at $\Gamma$ point, and the electrons on $E_-$ ($P_{x-iy,\uparrow}^+$ or $P_{x+iy,\downarrow}^+$) can be excited to $E_+$ ($P_{x-iy,\downarrow }^+$ or $P_{x+iy,\uparrow}^+$) by thermal activation or parallel magnetic field. In high temperature region, both levels are partially filled, so the superposition of up-spin and down-spin of the same orbit $P_{x+iy}^+$ (or $P_{x-iy}^+$) weakens the spin polarization along $z$-direction and the phenomena is like 2D GL and KLB theory. By contrast if $T$ is close to zero, the upper band is almost empty so its influence is negligible, and the electrons have robust spin polarization, making the Cooper pairing very difficult to break by the parallel field and leading to the upturn of $B_c$ in low temperature region.

\textit{Inversion-asymmetric Ising superconductivity.---}
	For in-plane inversion asymmetric systems, the inversion-asymmetric Ising superconductivity is caused by Zeeman-type SOI. Considering 2D hexagonal lattice as an example, the SOI serves as out-of-plane magnetic field which takes opposite value $B_{eff}\hat{z}, -B_{eff}\hat{z}$ at valleys $K$ and $K'$, and the spin-degeneracy of energy bands are lifted, resulting in double FSs with almost the same radius and shape around each valley [see Fig. \ref{Fig schematic}(d)]. Thus, electrons at $\bm{K}+\bm{k}_r$ and $-\bm{K}-\bm{k}_r$ ($r=1,2$ labels the two FSs) have opposite out-of-plane spin because of TRS [see Fig. \ref{Fig schematic}(c)], and intervalley Cooper pairs formed by those electrons are stable under an in-plane field much larger than $B_{p}$. When an in-plane magnetic field $B\hat{x}$ is present, the normal state Hamiltonian reads $H_{I}({k}) = \frac{\hbar^2k^2}{2m} -\beta_{so}\sigma_z\tau_z -\mu_BB\sigma_x$\cite{PhysRevLett.113.097001}, where $m$ is the effective mass, $\sigma$ and $\tau$ denotes the real spin subspace and valley subspace respectively. The relation between critical field $B_c$ and temperature $T$ can be solved within the WHH framework\cite{PhysRevX.8.021002}, and the in-plane critical field satisfies the equation:
    \begin{equation}\label{ia-eq}
			\ln\left(\frac{T_{c}}{T}\right) = \sum_{n=0}^{\infty}
			\frac{2\pi k_BT \mu_B^2 B_c^2}
			{ \omega_n\big[ \omega_n^2 + \frac{\beta_{so}^2}{\left( 1 + \hbar / 2\omega_n\tau_0 \right)^2} + \mu_B^2 B_c^2 \big] },
		\end{equation}
with digamma function $\psi(x)$, Matsubara frequencies $\omega_n=(2n+1)\pi k_BT$ and spin-independent scattering $\tau_0$. If the spin-independent scattering is weak, or the temperature is not too low, the simplified equation is $F(\tilde{\beta}_{so},t,b)=0$ in terms of dimensionless effective Zeeman-type SOI	$\tilde{\beta}_{so} = \frac{\beta_{so}}{k_BT_{c}+\hbar/(2\pi \tau_0)}$.
Instead of using Dyson equations technique to solve the inversion-asymmetric problem\cite{PhysRevLett.119.117001}, the WHH method here finishes the work after the fashion of the aforementioned inversion-symmetric Ising case, showing its convenience in both types of Ising superconductivity.
We note that when the Eq. (\ref{is-eq}) of inversion-symmetric returns to $F(\tilde{m}_1,t,b)=0$, the functional form looks like the inversion-asymmetric Ising case, but the new effective parameter $\tilde{m}_1$ is not the Zeeman-type SOI. This similarity in mathematical form suggests the universality of the $F(\tilde{m},t,b)$ function in various types of Ising superconductivity.

\begin{figure}
		\includegraphics[width=8.6cm]{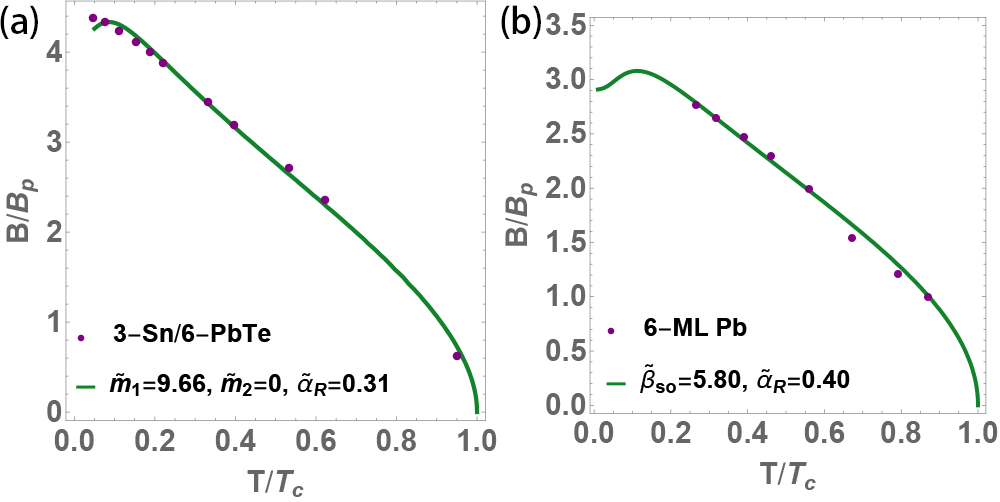}
		\caption{(Color online) Normalized in-plane critical field $B/B_{p}$ as a function of reduced temperature $T/T_{c}$ in different types of Ising superconducting systems. (a) Few-layer stanene 3-Sn/6-PbTe, $T_{c}=0.45 \mathrm{K}$ and $B_{p}=0.84 \mathrm{T}$, fitted by inversion-symmetric Ising theory, and the experimental data is from from Ref.\cite{falson2019type}. The fitting shows a weak Rashba parameter $\tilde{\alpha}_R\ll \tilde{m}_1$ for $k_1$ FS, and the $k_2$ FS has no effective field possibly because $k_2$ is too large. The BCS coupling constants are $\lambda_{11}=3, \lambda_{22}=1, \lambda_{12}=0.4$. (b) 6-monolayer (ML) Pb films, $T_{c}=6.00 \mathrm{K}$ and $B_{p}=14.7 \mathrm{T}$, fitted by inversion-asymmetric Ising theory, and the experimental data is from from Ref.\cite{PhysRevX.8.021002}. % Remark to Hongchao: the fitting and data in Fig.3A need to be changed to 3-Sn/6-PbTe for higher B/Bp and recheck the parameters after the revision. 
		}\label{fitting}
	\end{figure}

\textit{Discussions.---}
%In addition to the spin-orbit scattering captured by the KLB theory, other type of spin-orbit interaction may also influence the two types of Ising superconductivity.
There may be multiple types of SOI working simultaneously, including the Ising-inducing SOI and the Rashba SOI to affect the in-plane $B_c$ in experiments \cite{lu2015, saito2016}.
Considering the effect of Rashba SOI originating from the interface, both the inversion-symmetric formula Eq. (\ref{is-eq})(\ref{is-eqF}) and the inversion-asymmetric formula Eq. (\ref{ia-eq}) can be further modified to include the influence of Rashba SOI, and further be utilized to fitting the in-plane critical field $B_c$ of few-layer stanene and ultrathin crystalline Pb films, respectively\cite{supmatthis}.
In both case, the weak Rashba-type SOI tends to polarize the spin to the in-plane direction, making the Cooper pairs more susceptible to the in-plane magnetic field, and destructs the upturn at very low temperature.
If Rashba-type SOI is strong, the upturn in low temperature region will be completely destroyed.
We use both types of formulas for Ising superconductivity with dimensionless effective Rashba-type SOI $\tilde{\alpha}_R=\frac{\alpha_Rk_{F}/\sqrt{2}}{k_BT_{c}+\hbar/(2\pi \tau_0)}$ to fit the experimental data quantitatively [see Fig. \ref{fitting}], and the results give very weak Rashba-type SOI parameter $\tilde{\alpha}_R\ll \tilde{m}_1$ or $\tilde{\alpha}_R\ll \tilde{\beta}_{so}$.

\begin{figure}
		\includegraphics[width=0.45\textwidth]{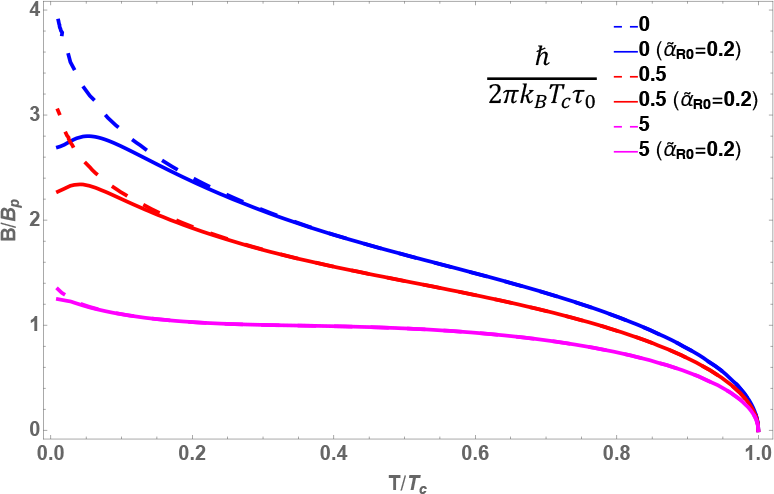}
		\caption{(Color online) In-plane critical field normalized by $B_{p}$ as a function of temperature normalized by $T_{c}$, for inversion-symmetric Ising formula Eq. (\ref{is-eq}).
		The solid lines are the $B_c-T$ relations with Rashba SOI, and the dashed lines are those without Rashba SOI.
		The SOI parameters are the same for all curves $m_{1,0}=4.6, m_{2,0}=0.5$ and $\tilde{\alpha}_{R0}=0.2$, while the values of $\tau_0$ are different. The BCS coupling constants are $\lambda_{11}=3, \lambda_{22}=1, \lambda_{12}=2$.} \label{tau}
	\end{figure}

In the above derivation, changing the disorder strength renormalizes every effective SOI parameter in the same way $\tilde{X}= \frac{X_0}{1+\hbar/(2\pi \tau_0 k_BT_{c})}$, where $X=m_1,m_2,\beta_{so},\alpha_R$ and $X_0$ denotes the original dimensionless SOI.
Therefore, the curves in Fig. \ref{tau} of smaller $\tau_0$ gives smaller $B_c(T)$ for any $T<T_c$ because of smaller $\tilde{m}_1,\tilde{m}_2$, meanwhile the effect of Rashba SOI is weaker.
The solid lines in Fig. \ref{tau} overlap more with the dashed ones for smaller $\tau_0$, showing the Rashba effect strongly restricted in a narrower low-temperature region as the system gets dirtier.

It should be noted that the Hamiltonian model Eq. (\ref{stanene model}) for inversion-symmetric Ising theory is block-diagonal at zero $B$, and the eigenstates at the $r$th FS can be written as $P_{x-iy,\uparrow}^+ + \eta P_{x+iy,\uparrow}^+$ and $P_{x+iy,\downarrow}^+ - \eta^* P_{x-iy,\downarrow}^+$, where $|\eta|=v k_r/2M_0+O(k_r^2)$.
If $k_r$ is close to zero, then the eigenstates can be approximated by $P_{x-iy,\uparrow}^+$ and $P_{x+iy,\downarrow}^+$, and the SOI can be viewed as a result of orbit-locked $\pm B_{eff}\hat{z}$ mentioned before [see Fig. \ref{Fig schematic}(d)]. If $k_r$ is too large, the coupling parameter $\eta$ cannot be neglected, which gives rise to smearing out of Ising pairing. In other words, the effect of the $\eta$ is similar to that of the Rashba SOI, and thus can also contribute to bend downward the critical field in the ultra-low temperature regime.
%Meanwhile, for inversion-asymmetric Ising system, although strict-diagonal at zero $B$, the inversion-asymmetric Ising model is only a good approximation near the valley points.
%For electrons on the FS of large $k_{F}$, Zeeman-type spin splitting along $z$-direction may no longer be dominant, making the in-plane critical field trivial.
%Thus, both types of Ising superconductivity may be fragile in systems with large Fermi momentum.

\textit{Summary.---} We propose the microscopic theory for the in-plane critical field of two-dimensional Ising superconducting systems, including systems with or without inversion symmetry breaking. In both systems, the intrinsic spin-orbit interaction polarizes the electron spin to the out-of-plane direction, which gives rise to large in-plane critical field surpassing the Pauli limit. Meanwhile, the critical field shows remarkable upturn near the zero temperature limit. The microscopic theory can quantitatively explain recent experimental results in stanene and Pb superconducting ultra-thin films.

\textit{Acknowledgements.---} H. Liu is grateful for J. Wang and Y. Liu for their insightful discussions throughout the previous collaborations on related works. This work was financially supported by the National Basic Research Program of China (Grants No. 2017YFA0303301, No. 2015CB921102), the National Natural Science Foundation of China (Grants No. 11674028, No. 11534001, No. 11504008), and the Fundamental Research Funds for the Central Universities.

\bibliography{Ising}

\end{document}

\title{Supplemental Material for ``Microscopic theory of in-plane critical field in two-dimensional Ising superconducting systems''}
\author{Hongchao Liu}
\affiliation{International Center for Quantum Materials, Peking University, Beijing 100871, China}
\author{Haiwen Liu}
% \email{haiwen.liu@bnu.edu.cn}
\affiliation{Center for Advanced Quantum Studies, Department of Physics, Beijing Normal University, Beijing 100875, China}
\author{Ding Zhang}
\affiliation{State Key Laboratory of Low-Dimensional Quantum Physics, Department of Physics, Tsinghua University, Beijing 100084, China}
\affiliation{Beijing Academy of Quantum Information Sciences, Beijing 100193, China}
\author{X. C. Xie}
\affiliation{International Center for Quantum Materials, Peking University, Beijing 100871, China}
\affiliation{Beijing Academy of Quantum Information Sciences, Beijing 100193, China}
\affiliation{CAS Center for Excellence in Topological Quantum Computation, University of Chinese Academy of Sciences, Beijing 100190, China}
\date{\today{}}
\maketitle\

\section{The normal state Hamiltonian of inversion-symmetric Ising superconducting system}

	We consider the typical system of inversion-symmetric Ising superconducting Stanene system with normal state Hamiltonian described by the Bernevig-Hughes-Zhang (BHZ) model \cite{PhysRevLett.111.136804}. The basis of the matrix is $(P_{x+iy,\uparrow}^+, P_{x-iy,\uparrow}^+, P_{x-iy,\downarrow}^+, P_{x+iy,\downarrow}^+ )$, where the superscript $(+,-)$ denotes the parity. 
		\begin{equation}\label{hamil1}\begin{aligned}
			H (\bm{k}) = & A k^2 + \left[ \begin{array}{c c} 
				H_+(\bm{k})   &  -\mu_\mathrm{B}B\sigma_x  \\ -\mu_\mathrm{B}B\sigma_x  & H_- (\bm{k})
				\end{array} \right] ,  \\
			= & A k^2 + (M_0-M_1k^2)\sigma_z +vk (\cos\theta\sigma_x\tau_z + \sin\theta\sigma_y ) - \mu_\mathrm{B}B\sigma_x\tau_x \\
			H_\pm(\bm{k}) = & \left[ \begin{array}{c c} 
				M_0-M_1k^2   & \quad v (\pm k_x-i k_y)  \\ 
				v (\pm k_x+i k_y)  & \quad -M_0+M_1k^2
				\end{array} \right].
		\end{aligned}\end{equation}
	The in-plane magnetic field induces spin splitting in $x$-direction: 
	\begin{equation}
		-\mu_\mathrm{B}B(P_{x+iy,\uparrow}^{+\dagger}P_{x+iy,\downarrow}^+ +P_{x-iy,\uparrow}^{+\dagger}P_{x-iy,\downarrow}^+ +h.c.)
		= \left[ \begin{array}{c c} 
			0 &  -\mu_\mathrm{B}B\sigma_x  \\ 
			-\mu_\mathrm{B}B\sigma_x  & 0 \end{array} \right].
	\end{equation} 
	The Zeeman term doesn't have TRS and changes sign under time reversal
	\begin{align}
		% \hat{T}P_{x+iy,\uparrow}^{+\dagger}P_{x+iy,\downarrow}^+\hat{T}^{-1} = & -P_{x-iy,\downarrow}^{+\dagger}P_{x-iy,\uparrow}^+ \nonumber \\
		\hat{T}\left[ \begin{array}{c c} 
			0 &  -\sigma_x  \\ -\sigma_x  & 0
			\end{array} \right]\hat{T}^{-1} 
		= &\left[ \begin{array}{c c} 
			0 &  \sigma_x  \\ \sigma_x  & 0
			\end{array} \right]. 
	\end{align}

	The energy band $E(k)$ of Eq. (\ref{hamil1}) at $B=0$ (the four eigenvalues are doubly degenerate)
		\begin{equation}\label{Ek}
		E_{1,2}(k)=Ak^2 \pm \sqrt{( M_0-M_1k^2 )^2+v^2k^2},
		\end{equation}
	and we assume $\eta = - \dfrac{vk e^{-i\theta}}{\sqrt{( M_0-M_1k^2 )^2+v^2k^2} + M_0-M_1k^2 } $, then the eigenstates of $E_1(k)$ and $E_2(k)$ read 
		\begin{equation}
		\left[ \begin{array}{l}
			0 \\ 0 \\ 1 \\ \eta  
		\end{array} \right],
		\left[ \begin{array}{l}
			1 \\ -\eta^* \\ 0 \\ 0
		\end{array} \right], \quad \mathrm{and} \quad
		\left[ \begin{array}{l}
			\eta \\ 1 \\ 0 \\ 0 
		\end{array} \right],
		\left[ \begin{array}{l}
			0 \\ 0 \\ -\eta^* \\ 1  
		\end{array} \right],
		\end{equation}
	The $E_2(k_r) -E_F=0$ points are (assumed to be) two circles in the $xy$ plane with ``Fermi vectors'' $k_1$ and $k_2$ ($0<k_1<k_2$), forming two Fermi surfaces (FSs) denoted by FS index $r=1,2$. We assume FS $r$ has an approximate Hamiltonian $H_r=A_r(k^2-k_r^2)+H(\bm{k}_r) $ in which the $H(\bm{k}) $ matrix is ``fixed'' at $k_r$.
		\begin{equation}
		H_r = A_r (k^2-k_r^2) + Ak_r^2+(M_0-M_1k_r^2)\sigma_z +vk_r (\cos\theta\sigma_x\tau_z + \sin\theta\sigma_y ) - \mu_\mathrm{B}B\sigma_x\tau_x \\,
		\end{equation}
where $A_1,A_2$ are ``effective mass'' to make the slope of $E_i(k)$ curve at $k_r$ equal to those of Eq. (\ref{Ek}). We assume the interband scattering can be neglected, and treat $H_1, H_2$ separately in the next section.

\section{Critical field for one Fermi surface}

	Based on the Gor'kov Green's function technique \cite{abrikosov1962}, we consider only one Fermi surface present, e.g. band 1. The spin-independent scattering disorder is denoted by mean free time $\frac{\hbar}{\tau_0}=2\pi n_i N(0) u_1^2$, where $n_i$ is the density of impurities. The Green function under influence of non-magnetic disorder scattering is
		\begin{equation}\label{gk}
		G_\omega^n(\bm{k})= \frac{1}{ i \tilde{\omega} - \left( A_1 (k^2-k_1^2) + H(\bm{k}_1) \right) },
		\end{equation}
where $\tilde{\omega}=\omega+ \frac{\hbar\sgn (\omega)}{2\tau_0},\,\omega = (2n+1)\pi k_\mathrm{B}T$ and $H(\bm{k}, B)$ is from Eq. (\ref{hamil1}). Under time reversal only the $B$ term change sign
		\begin{equation}\label{g-k}
		\left.G_{-\omega}^n(\bm{-k})\right|_{\sigma\rightarrow -\sigma}= \frac{1}{ -i \tilde{\omega} - \left( A_1 (k^2-k_1^2) + \left.H(\bm{k}_1)\right|_{B \rightarrow -B} \right) }. 
		\end{equation}

	Average the gap function over impurity configurations \cite{PhysRev.147.295} 
		\begin{equation}\label{Gap function 147}\begin{aligned}
		& \bar{\Delta}_\omega(\bm{r}-\bm{r}') = \delta^2(\bm{r}-\bm{r}')\Delta + \int d^2\bm{r}_1 d^2\bm{r}_2 
		\left\langle V(\bm{r},\bm{r}_1)F_\omega(\bm{r}_1-\bm{r}_2)V(\bm{r}',\bm{r}_2)\right\rangle \\
		= & \delta^2(\bm{r}-\bm{r}')\Delta + u_1^2 F_\omega(\bm{r}-\bm{r}') 
		\int \frac{d^2\bm{p} d^2\bm{q}}{(2\pi)^4} 
		\exp \left(i\bm{p}\bm{r}+i\bm{q}\bm{r}' \right) 
		\sum_{i,j} \left\langle \exp\left(-i\bm{p}\bm{R}_i-i\bm{q}\bm{R}_j \right) \right\rangle 
		\\
		= & \delta^2(\bm{r}-\bm{r}') \left(\Delta + n_i u_1^2 F_\omega(0) \right) + n_i^2 u_1^2 F_\omega(\bm{r}-\bm{r}'),
		\end{aligned}\end{equation}
	with $V(\bm{r},\bm{r}') $ denoting the non-magnetic impurity scattering and $\Delta$ denoting the superconducting gap. 
		\begin{equation}\label{V at zero u2}
			V(\bm{r},\bm{r}')=\delta^2(\bm{r}-\bm{r}') \sum_i \int \frac{d^2\bm{p}}{(2\pi)^2} 
			u_1 \exp \left(i\bm{p}\cdot(\bm{r}-\bm{R}_i) \right)
		\end{equation}

	The anomalous Green's function $F_\omega(\bm{r}-\bm{r}')$ is defined by:
		\begin{equation}\begin{aligned}
			F_\omega(\bm{r}-\bm{r}') = & \int d^2\bm{r}_1 d^2\bm{r}_2 G_\omega^n(\bm{r}-\bm{r}_1) \bar{\Delta}_\omega(\bm{r}_1-\bm{r}_2) \left.G_{-\omega}^n(\bm{r}'-\bm{r}_2)\right|_{\sigma\rightarrow -\sigma} \\ 
			\approx & \int d^2\bm{r}_1 G_\omega^n(\bm{r}-\bm{r}_1) \left(\Delta + n_i u_1^2 F_\omega(0) \right) \left.G_{-\omega}^n(\bm{r}'-\bm{r}_1)\right|_{\sigma\rightarrow -\sigma} \\
			F_\omega(\bm{k}) = & F_\omega^0(\bm{k}) + n_i u_1^2 G_\omega^n(\bm{k}) F_\omega(0) \left.G_{-\omega}^n(\bm{-k})\right|_{\sigma\rightarrow -\sigma} 
		\end{aligned}\end{equation}
	And the transition temperature $T $ obeys
		\begin{equation}
		\ln\left(\frac{T_\mathrm{c}}{T}\right) = k_\mathrm{B} T \sum_{n=-\infty}^{\infty}\left(\frac{\pi}{|\omega|}- \frac{1}{4} \Tr \left(\frac{F_\omega(0)}{N(0)\Delta}\right) \right)
		\end{equation}
		\begin{equation}
		F_\omega(0) = \int d^2\bm{k} F_\omega(\bm{k}) = \frac{1}{2|A_1|} \int d\xi d\theta F_\omega(\bm{k}) \equiv \frac{\Delta N(0)}{k_\mathrm{B}T} \overline{S_\omega}
		\end{equation}
	where $\xi=|A_1| k^2$, $N(0)=\frac{\pi}{|A_1|}$ \label{n0} is the density of states at Fermi level, and $\overline{f(\theta)}\equiv \int\frac{d\theta }{2\pi} f(\theta)$ with $S_\omega(\theta)$ the dimensionless integral kernel function \cite{PhysRevX.8.021002}
		\begin{equation}\label{bare full}
			S_\omega(\theta) = \frac{k_\mathrm{B}T}{\Delta} \int d\xi F_\omega(\bm{k}) 
			\ = S_\omega^0(\theta) + n_i u_1^2 N(0) \int d\xi  \left( G_\omega^n(\bm{k}) \overline{S_\omega} \left.G_{-\omega}^n(\bm{-k})\right|_{\sigma\rightarrow -\sigma} \right) 
		\end{equation}
	The bare anomalous Green’s function is
		\begin{equation}
			F_\omega^0(\bm{r}-\bm{r}') = \int d^2\bm{r}_1 G_\omega^n(\bm{r}-\bm{r}_1) \Delta \left.G_{-\omega}^n(\bm{r}'-\bm{r}_1)\right|_{\sigma\rightarrow -\sigma} 
		\end{equation}
	Introduce an integral identity to compute the bare integral kernel $S_\omega^0(\theta)$ from normal Green's function $G_\omega^n(\bm{k})$
		{\small \begin{equation}\begin{aligned}
			& \int\limits_{-\infty}^\infty dx (x-i + A'\sigma_z + B'\sigma_x\tau_z + C' \sigma_y + D'\sigma_x\tau_x+ E'\sigma_x\tau_x + F'\sigma_x\tau_y)^{-1} \cdot(x+i+ A'\sigma_z \\
			& + B'\sigma_x\tau_z + C' \sigma_y - D'\sigma_x\tau_x+ E'\sigma_x\tau_x + F'\sigma_x\tau_y)^{-1}  \\
			= & \pi \bigg( 1 + \frac{i D'}{1+A'^2+B'^2+C'^2+E'^2+F'^2} \cdot \Big( (1+E'^2)\sigma_x\tau_x + A'\sigma_y\tau_x + B'\tau_y - C'\sigma_z\tau_x-F'\tau_z \\
			& + E'(A'\sigma_z + B'\sigma_x\tau_z + C'\sigma_y + F'\sigma_x\tau_y) \Big) \bigg)^{-1} 
		\end{aligned}\end{equation} }
	Then $S_\omega^0(\theta)$ is 
		{\small \begin{equation}\begin{aligned}\label{kernel0}
		\left[S_\omega^0(\theta)\right]^{-1} = & \left(k_\mathrm{B}T \int d\xi G_\omega^n(\bm{k}) \left.G_{-\omega}^n(\bm{-k})\right|_{\sigma\rightarrow -\sigma}\right)^{-1} \\
		= & \frac{|\tilde{\omega}|}{\pi k_\mathrm{B}T} \left( 1 + \frac{-i \mu_\mathrm{B} B \big( \tilde{\omega}\sigma_x\tau_x + (M_0-M_1k_1^2 )\sigma_y\tau_x  + vk_{1x} \tau_y - vk_{1y}\sigma_z\tau_x \big) }{ \tilde{\omega}^2 + (M_0-M_1k_1^2 )^2 +v^2k_1^2} \right)
		\end{aligned}\end{equation} }
	Considering the disorder influence, the full integral kernel $S_\omega(\theta)$ is given be Eq. (\ref{bare full}): 
		{\small \begin{equation}\begin{aligned}\label{S}
			& \left[S_\omega(\theta)\right]^{-1} = \left[S_\omega^0(\theta)\right]^{-1}-\frac{\hbar}{2\pi k_\mathrm{B}T\tau_0} \\
			= & \frac{ |\omega| }{\pi k_\mathrm{B}T} \left( 1 + (1+\frac{\hbar}{2|\omega|\tau_0})\frac{-i \mu_\mathrm{B} B \big( \tilde{\omega}\sigma_x\tau_x + (M_0-M_1k_1^2 )\sigma_y\tau_x  + vk_{1x} \tau_y - vk_{1y}\sigma_z\tau_x \big) }{ \tilde{\omega}^2 + (M_0-M_1k_1^2 )^2 +v^2k_1^2} \right)
		\end{aligned}\end{equation} }
	The relation between the upper critical field $B_c(T)$ and temperature $T$ is \cite{PhysRev.147.295}
		\begin{equation}\label{BT0}
			\ln\left(\frac{T_\mathrm{c}}{T}\right) = k_\mathrm{B}T \sum_{n=-\infty}^{\infty}\left(\frac{\pi}{|\omega|}-\frac{1}{4k_\mathrm{B} T} \Tr \overline{S_\omega} \right) 
		\end{equation}
	
	Then the relation between $B$ and $T$ is in the following form
		\begin{equation}\label{BT1 not summed}
			\ln\left(\frac{T_\mathrm{c}}{T}\right) = \pi k_\mathrm{B}T \sum_{n=-\infty}^{\infty} \frac{1}{|\omega|} \frac{\mu_\mathrm{B}^2 B^2}{ \omega^2 + \frac{ (M_0-M_1k_1^2)^2 + v^2k_1^2}{\left( 1 + \hbar / 2|\omega|\tau_0 \right)^2} + \mu_\mathrm{B}^2 B^2}  .
		\end{equation}
	We assume the approximation $ 1 + \frac{\hbar}{2|\omega|\tau_0} \approx 1 + \frac{\hbar}{2\pi k_\mathrm{B}T_\mathrm{c}\tau_0}$ and introduce new dimensionless parameters
		\begin{equation}\label{para OB}
		\tilde{m}=\frac{ \sqrt{ (M_0-M_1k_1^2)^2 + v ^2k_1^2 }}{k_\mathrm{B}T_\mathrm{c}+\hbar/(2\pi \tau_0) }, \quad t=\frac{T}{T_\mathrm{c}}, \quad b=\frac{\mu_\mathrm{B}B}{k_\mathrm{B}T_\mathrm{c}}, 
		\end{equation}
	By performing the summation
	{\small	
		\begin{equation}\label{summation1} \begin{aligned}
			& \sum_{n=-\infty}^{\infty} \frac{\pi k_\mathrm{B}T }{|\omega|}  \frac{\tilde{m}^2 +b^2}{ (\frac{\omega}{k_\mathrm{B}T_\mathrm{c}})^2 + \tilde{m}^2 +b^2}   = \sum_{n=0}^{\infty} \left( \frac{1}{n+\frac{1}{2}} -   \frac{ n+\frac{1}{2} }{ (n+\frac{1}{2})^2 + \frac{\tilde{m}^2 +b^2}{(2\pi t)^2} } \right)   \\
			= & \Real \sum_{n=0}^{\infty} \left( \frac{1}{n+\frac{1}{2}} - \frac{1}{ n+\frac{1}{2} + \frac{i \sqrt{\tilde{m}^2 +b^2}}{2\pi t} } \right) = \Real \psi\left(\frac{1}{2}+\frac{i \sqrt{\tilde{m}^2 +b^2}}{2\pi t}\right) - \psi\left(\frac{1}{2}\right) 
		\end{aligned}\end{equation} 
	}
	we get the upper critical field $B_c(T)$ 
		\begin{equation}\label{BT1}
			\ln t + \frac{b^2}{\tilde{m}^2 +b^2} \bigg[ \Real \psi \bigg(\frac{1}{2}+\frac{i\sqrt{\tilde{m}^2 +b^2}}{2\pi t}\bigg)  -\psi\bigg(\frac{1}{2}\bigg) \bigg] =0,
		\end{equation}
	with $\psi(z)$ denoting the digamma function. The effect of the original Hamiltonian model ($M_0,M_1,v,k_1$) and disorder scattering ($\tau_0$) are all summarized into one parameter $\tilde{m}$. 

	If we only consider the 2nd FS, the result is virtually the same, except $k_1$ is replaced by $k_2$ in Eqs. (\ref{gk})-(\ref{BT1}). 

	{\small 
	\textbf{With Rashba-type SOI}

	If the Rashba-type SOI is present, more complicated formulae can be derived. The Rashba term in the basis $(P_{x+iy,\uparrow}^+ , P_{x-iy,\uparrow}^+ , P_{x-iy,\downarrow}^+ , P_{x+iy,\downarrow}^+ )$ has TRS 
	\begin{equation}\label{HR}
		H_\mathrm{R} = -\alpha_\mathrm{R}\left[(k_y+ik_x)(P_{x+iy,\uparrow}^{+\dagger}P_{x+iy,\downarrow}^+ +P_{x-iy,\uparrow}^{+\dagger}P_{x-iy,\downarrow}^+) +h.c.\right] 
		= -\alpha_\mathrm{R}(k_y\tau_x-k_x\tau_y)\sigma_x  
	\end{equation}
	The approximate Hamiltonian model is $H_{r}+H_R(k_r)$. We can repeat the one-band analysis in Section II with this new Hamiltonian, and get 
		\begin{equation}\begin{aligned}\label{kernel0R}
		\left[S_\omega(\theta)\right]^{-1} 
		= & \frac{|\omega|}{\pi k_\mathrm{B}T} \bigg( 1 + (1+\frac{\hbar}{2|\omega|\tau_0})\frac{-i \mu_\mathrm{B} B }{ \tilde{\omega}^2+ (M_0-M_1k_1^2 )^2 +v^2k_1^2 +\alpha_\mathrm{R}^2k_1^2} \\
		& \cdot \Big( (\tilde{\omega} + \frac{\alpha_\mathrm{R}^2k_{1y}^2}{\tilde{\omega}})\sigma_x\tau_x + (M_0-M_1k_1^2 )\sigma_y\tau_x + vk_{1x} \tau_y - vk_{1y}\sigma_z\tau_x -\alpha_\mathrm{R}k_{1x} \tau_z  \\
		& - \frac{\alpha_\mathrm{R}k_{1y}}{\tilde{\omega}} \big( (M_0-M_1k_1^2 )\sigma_z +vk_{1x} \sigma_x\tau_z + vk_{1y}\sigma_y+\alpha_\mathrm{R}k_{1x} \sigma_x\tau_y \big) \Big) \bigg)
		\end{aligned}\end{equation}
	where $\tilde{\omega}=\omega+ \frac{\hbar\sgn (\omega)}{2\tau_0},\,\omega = (2n+1)\pi k_\mathrm{B}T$. This equation falls back to Eq. (\ref{S}) at $\alpha_\mathrm{R}=0$. The one-band $T-B$ relation is 
		{\small \begin{equation}\label{onebandR}
			\ln \left(\frac{T_\mathrm{c}}{T}\right) = \sum_{n=-\infty}^{\infty} \frac{ \pi k_\mathrm{B}T }{|\omega|} 		\overline{ 		\left[ \frac{\mu_\mathrm{B}^2 B^2(1+ \frac{ \alpha_\mathrm{R}^2k_{1}^2\sin^2\theta }{\tilde{\omega}^2} )}  { \omega^2 + \frac{ (M_0-M_1k_1^2)^2 + v^2k_1^2 + \alpha_\mathrm{R}^2k_1^2}{\left( 1 + \hbar / 2|\omega|\tau_0 \right)^2} + \mu_\mathrm{B}^2 B^2(1+ \frac{ \alpha_\mathrm{R}^2k_{1}^2\sin^2\theta }{\tilde{\omega}^2} )} \right] 	}
		\end{equation} }
	We assume the approximation $1 + \hbar / 2|\omega|\tau_0 \approx 1 + \hbar / 2\pi k_\mathrm{B}T_\mathrm{c}\tau_0$, and introduce new dimensionless parameters (FS index $r=1,2$)
	\begin{equation}\label{para_onebandR}
		\tilde{m}=\frac{ \sqrt{ (M_0-M_1k_1^2)^2 + v ^2k_1^2 }}{k_\mathrm{B}T_\mathrm{c}+\hbar/(2\pi \tau_0) }, 
		\quad \tilde{\alpha}_\mathrm{R}=\frac{1}{\sqrt{2}}\frac{\alpha_\mathrm{R}k_1}{k_\mathrm{B}T_\mathrm{c}+\hbar/(2\pi \tau_0) },
		\quad t=\frac{T}{T_\mathrm{c}}, 
		\quad b=\frac{\mu_\mathrm{B}B}{k_\mathrm{B}T_\mathrm{c}}, 
	\end{equation}
	We also assume the angular average can be calculated separately, then finish a summation much more complicated than Eq. (\ref{summation1})
	\begin{equation}\begin{aligned}\label{tempb13}
		-\ln t  \approx & \sum_{n=-\infty}^{\infty} \frac{ \pi k_\mathrm{B}T }{|\omega|} \frac{b^2 (\frac{\omega}{k_\mathrm{B}T_\mathrm{c}})^2 + b^2 \tilde{\alpha}_\mathrm{R}^2 }{ (\frac{\omega}{k_\mathrm{B}T_\mathrm{c}})^2 \big((\frac{\omega}{k_\mathrm{B}T_\mathrm{c}})^2 + \tilde{m}^2 +2\tilde{\alpha}_\mathrm{R}^2 +b^2 \big) + b^2 \tilde{\alpha}_\mathrm{R}^2 } \\
		= & \sum_{n=0}^{\infty} \frac{1}{n+\frac{1}{2}} \frac{ \frac{b^2}{(2\pi t)^2} (n+\frac{1}{2})^2 + \frac{b^2 \tilde{\alpha}_\mathrm{R}^2 }{(2\pi t)^4} }{ (n+\frac{1}{2})^2 \big( (n+\frac{1}{2})^2 + \frac{\tilde{m}^2 +2\tilde{\alpha}_\mathrm{R}^2 +b^2}{(2\pi t)^2} \big) + \frac{b^2 \tilde{\alpha}_\mathrm{R}^2 }{(2\pi t)^4} } \\
		= & \sum_{n=0}^{\infty} \left( \frac{1}{n+\frac{1}{2}} - \frac{ (n+\frac{1}{2}) [ (n+\frac{1}{2})^2 + \frac{\tilde{m}^2 +2\tilde{\alpha}_\mathrm{R}^2}{(2\pi t)^2} ] }	{ (n+\frac{1}{2})^4 + \frac{\tilde{m}^2 +2\tilde{\alpha}_\mathrm{R}^2 +b^2}{(2\pi t)^2}(n+\frac{1}{2})^2  + \frac{b^2 \tilde{\alpha}_\mathrm{R}^2 }{(2\pi t)^4} } \right) \\
		= & \sum_{n=0}^{\infty} \left( \frac{1}{n+\frac{1}{2}} - (n+\frac{1}{2}) \Big( \frac{ \frac{1}{2}(1-A) }{(n+\frac{1}{2})^2+(\frac{\rho_{+}}{2\pi t})^2} + \frac{ \frac{1}{2}(1+A) }{(n+\frac{1}{2})^2+(\frac{\rho_{-}}{2\pi t})^2} \Big) \right) \\
		= & \frac{1-A}{2} \sum_{n=0}^{\infty} \left( \frac{1}{n+\frac{1}{2}} - \frac{ n+\frac{1}{2} }{(n+\frac{1}{2})^2+(\frac{\rho_{+}}{2\pi t})^2} \right) 		+ \frac{1+A}{2} \sum_{n=0}^{\infty} \left( \frac{1}{n+\frac{1}{2}} - \frac{ n+\frac{1}{2} }{(n+\frac{1}{2})^2+(\frac{\rho_{-}}{2\pi t})^2}  \right) \\
		= & \frac{1-A}{2} \Real \psi\left(\frac{1}{2}+\frac{i\rho_{+}}{2\pi t}\right) 		+ \frac{1+A}{2} \Real \psi\left(\frac{1}{2}+\frac{i\rho_{-}}{2\pi t}\right) - \psi\left(\frac{1}{2}\right) 
	\end{aligned}\end{equation}

	where 
	\begin{equation}\label{para_onebandR_2}
		2\rho_{\pm} = \sqrt{\tilde{m}^2 +\tilde{\alpha}_\mathrm{R}^2 +(\tilde{\alpha}_\mathrm{R} +b)^2} \pm \sqrt{\tilde{m}^2 +\tilde{\alpha}_\mathrm{R}^2 +(\tilde{\alpha}_\mathrm{R} -b)^2}, \quad A= \frac{\tilde{m}^2 +2\tilde{\alpha}_\mathrm{R}^2 -b^2}{\rho_{+}^2-\rho_{-}^2},
	\end{equation} 
	This form is similar to the inversion-asymmetric Ising case with Rashba-type SOI in \cite{PhysRevX.8.021002}.
	}

\section{Extension to two-band inversion-symmetric Ising superconducting system}
	The notations in this section bring back the band index $r=1,2$, and the $T_\mathrm{c}$ in last section has to become $T_{\mathrm{c}1}$, and $T_\mathrm{c}$ in this section is the two-band critical temperature. The form of Eq. (\ref{para OB}) is almost unchanged, but the $T_\mathrm{c}$ is different, so we write again 
	\begin{equation}
		\tilde{m}\rightarrow\tilde{m}_r=\frac{ \sqrt{ (M_0-M_1k_r^2)^2 + v^2k_r^2 }}{ k_\mathrm{B}T_\mathrm{c}+\hbar/(2\pi \tau_0) }, \quad t=\frac{T}{T_\mathrm{c}}, \quad b=\frac{\mu_\mathrm{B}B}{k_\mathrm{B}T_\mathrm{c}},
	\end{equation}

	We can use the above one-band theory in separate bands, and then get them together with two-band Usedal equation \cite{PhysRevB.67.184515}. The Eq. (\ref{BT1}) for only one band can be rewritten for each band:
		\begin{equation}
			1 = \lambda_{11} (l-U(\tilde{m}_1,t,b)),\quad 1 = \lambda_{22} (l-U(\tilde{m}_2,t,b)),
		\end{equation}
		\begin{equation}
			U(\tilde{m}_r,t,b) = \frac{b^2}{\tilde{m}_r^2 +b^2} \bigg[ \Real \psi \bigg(\frac{1}{2}+\frac{i\sqrt{\tilde{m}_r^2 +b^2}}{2\pi t}\bigg) -\psi\bigg(\frac{1}{2}\bigg) \bigg] 
		\end{equation}
	where
		\begin{equation}
			l=\ln\frac{2\gamma \omega_D}{\pi k_\mathrm{B}T}, \quad \frac{1}{\lambda_{rr}} = \ln\frac{2\gamma \omega_D}{\pi k_\mathrm{B}T_{\mathrm{c}r}}, 
		\end{equation}
	are dimensionless, $\ln \gamma=0.577$ is the Euler constant, $\lambda_{rr'}$ are BCS superconducting coupling constants. The form of $U_r$ is dimensionless and independent of $T_{\mathrm{c}}$ (we assume $ \frac{\hbar}{2\pi k_\mathrm{B}\tau_0} \ll T_{\mathrm{c}r}, T_\mathrm{c}$). Here the diagonal terms $\lambda_{11}$ and $\lambda_{22}$ quantify the intraband superconducting coupling, and off-diagonal terms $\lambda_{12}$ and $\lambda_{21}$ describe the interband coupling.

	%$\lambda_{rr'}=\lambda_{rr'}^{(ep)}-\mu_{rr'} $ are BCS superconducting coupling constants, $\lambda_{rr'}^{(ep)}$ are electron-phonon constants, $\mu_{rr'}$ is the matrix of the Coulomb pseudopotential.
	
	Then we assume the equations in our case are 
		\begin{equation}\label{Delta1 Delta2}
			\begin{aligned}
			\tilde{\Delta}_1 = & \lambda_{11}(l-U(\tilde{m}_1,t,b)) \tilde{\Delta}_1 + \lambda_{12}(l-U(\tilde{m}_2,t,b)) \tilde{\Delta}_2 \\
			\tilde{\Delta}_2 = & \lambda_{22}(l-U(\tilde{m}_2,t,b)) \tilde{\Delta}_2 + \lambda_{21}(l-U(\tilde{m}_1,t,b)) \tilde{\Delta}_1
			\end{aligned}
		\end{equation}
	The solvability condition of Eq. (\ref{Delta1 Delta2}) gives the equation for $B_{c2}$ 
		\begin{equation}\label{two-band}
			\frac{2w}{\lambda_0} F(\tilde{m}_1,t,b) F(\tilde{m}_2,t,b) + \bigg(1+\frac{\lambda_-}{\lambda_0}\bigg) F(\tilde{m}_1,t,b) + \bigg(1-\frac{\lambda_-}{\lambda_0}\bigg) F(\tilde{m}_2,t,b) =0
		\end{equation}
	where $F(\tilde{m}_r,t,b)= \ln t + U(\tilde{m}_r,t,b)$,
		\begin{equation}
			w = \lambda_{11}\lambda_{22}-\lambda_{12}\lambda_{21},\quad \lambda_\pm = \lambda_{11}\pm\lambda_{22},\quad \lambda_0=\sqrt{\lambda_-^2+4\lambda_{12}\lambda_{21}}.
		\end{equation}
	$T_\mathrm{c}$ is defined as \cite{PhysRevB.67.184515}
		\begin{equation}\label{Tc for two-band}
			T_\mathrm{c}=\frac{2\gamma\omega_D}{\pi}\exp \left(-\frac{\lambda_+-\lambda_0}{2 w}\right).
		\end{equation}
	Eq. (\ref{two-band}) gives $t=1$ at $B=0$, showing $T_\mathrm{c}$ is the critical temperature at zero field in two band case.

	{\small 
	\textbf{With Rashba-type SOI}
	
	If the Rashba-type SOI is present Eq. (\ref{two-band}) becomes
	\begin{equation}
		\frac{2w}{\lambda_0} F_{1}^\mathrm{R} F_{2}^\mathrm{R} + \bigg(1+\frac{\lambda_-}{\lambda_0}\bigg) F_{1}^\mathrm{R} + \bigg(1-\frac{\lambda_-}{\lambda_0}\bigg) F_{2}^\mathrm{R} = 0
	\end{equation}
	where
	\begin{equation}
		F_{r}^\mathrm{R} = \ln t + \frac{1- A_r}{2} \Real \psi\left(\frac{1}{2}+\frac{i\rho_{r+}}{2\pi t}\right) + \frac{1+ A_r}{2} \Real \psi\left(\frac{1}{2}+\frac{i\rho_{r-}}{2\pi t}\right) - \psi\left(\frac{1}{2}\right) 
	\end{equation}
	and subscript $r=1,2$ labels the index of the FS. }

\section{Limiting behavior near $T_{c}$ and comparison to 2D FFLO state in the low-temperature regime}
	
	We write the temperature and the field in dimensionless form $t=\frac{T}{T_\mathrm{c}}, \ b=\frac{\mu_\mathrm{B}B}{k_\mathrm{B}T_\mathrm{c}}$. In the vicinity of $T_c$, $t\rightarrow 1$, and Ising theories gives asymptotic behavior $b\propto \sqrt{1-t}$, same as 2D Ginzburg–Landau (2D GL) and Klemm-Luther-Beasley (KLB) theory.
	\begin{enumerate}
	\item In 2D GL theory \cite{tinkham1996}, equation of the critical field in dimensionless form
	\begin{equation}
		b = \frac{\mu_\mathrm{B}}{k_\mathrm{B}T_\mathrm{c}}\frac{\Phi_0 \sqrt{12}}{2\pi \xi_{GL}(0)d_{sc}} \sqrt{1-t}
	\end{equation}
	where $\Phi_0$ is the flux quantum, $\xi_{GL}(0)$ is the GL coherence length at $T = 0 $K, and $d_{sc}$ is the effective thickness of superconductivity.
	\item In KLB theory \cite{PhysRevB.12.877}, equation of the critical field in dimensionless form
	\begin{equation}
		\ln t + \psi \bigg(\frac{1}{2}+\frac{b^2}{4\pi at} \bigg)-\psi\bigg(\frac{1}{2}\bigg)=0, \quad a= \frac{\hbar}{3k_\mathrm{B}T_\mathrm{c}\tau_\mathrm{so}} 
	\end{equation}
	At the vicinity of critical temperature, $\ln t \approx t-1,\, b^2/2a\ll 1,\, \psi(\frac{1}{2}+\frac{b^2}{4\pi a})-\psi(\frac{1}{2}) = \psi'(\frac{1}{2})\frac{b^2}{4\pi a} = \frac{\pi^2}{2}\frac{b^2}{4\pi a}$.
	\begin{equation}
		b = \sqrt{\frac{8a}{\pi}(1-t)}
	\end{equation}
	\item In inversion-asymmetric Ising theory,
	\begin{equation}
		\ln t + \frac{b^2}{\beta^2+b^2} \Real\bigg[ \psi\bigg(\frac{1}{2}+\frac{i\sqrt{\beta^2+b^2}}{2\pi t} \bigg) - \psi\bigg( \frac{1}{2}\bigg) \bigg] = 0
	\end{equation}
	At the vicinity of critical temperature, $\ln t \approx t-1, b\ll 1$,
	\begin{equation}
		b = \sqrt{ \frac{ \beta^2 (1-t) }{ \Real \psi\bigg(\frac{1}{2}+\frac{i \beta }{2\pi } \bigg) - \psi\bigg( \frac{1}{2}\bigg) } } 
	\end{equation}
	\item In inversion-symmetric Ising theory, at the vicinity of critical temperature, $\ln t \approx t-1, b\ll 1$,
	\begin{align}
		b = & \sqrt{\frac{2(1-t)}{a_1A_1+a_2A_2}} \\
		A_r = & \frac{1}{\tilde{m}_r^2} \bigg[ \Real \psi\bigg(\frac{1}{2}+\frac{i \tilde{m}_r }{2\pi } \bigg) - \psi\bigg( \frac{1}{2}\bigg) \bigg] \nonumber \\
		a_1 = & 1+\frac{\lambda_-}{\lambda_0}, \quad a_2 = 1-\frac{\lambda_-}{\lambda_0} \nonumber 
	\end{align}
	If $a_1\gg a_2$ or $\tilde{m}_1=\tilde{m}_2$, this fall back to one-band case.
	\end{enumerate}
	
	{\small 
	\textbf{Comparison to 2D FFLO state}
	
	To the our knowledge, the upturn feature was only seen previously in FFLO superconductors\cite{Shimahara2007,FFLOReview}. Here, we further compare the characteristic features of the FFLO state with the Ising superconductivity, and shows the remarkable difference between these two states. Firstly, the impurity scattering can destroy the possible FFLO states in dirty superconductor with $l\ll\xi$. In contrast, the impurity scattering renormalizes the effective spin splitting, as shown in Fig.4 of the main text. Moreover, for practical reason, we can provide tentative fittings of experimental data with the 2D FFLO Bc formula \cite{PhysRevB.50.12760}, as shown in Fig.S1. The quantitative comparison of 2D FFLO curve, the Ising paring curve, and the experimental data demonstate the remarkable differences between the former two theoretical models. Lastly, we want to mention that the Rashba SOC effect have opposite effect on the 2D FFLO state and the type-II Ising pairing. As shown in ref.\cite{doi:10.7566/JPSJ.86.083701}, the Rashba SOC prominently enhances Bc2 in 2D FFLO states. To the contrary, the Rashba SOC weakens the out-of-plane alignment of spin, and smears out the upturn feature in type-II Ising superconductivity, as shown in the comparison plots in Fig. 2b of the main text.
	
	\begin{figure}
		\includegraphics[width=8.6cm]{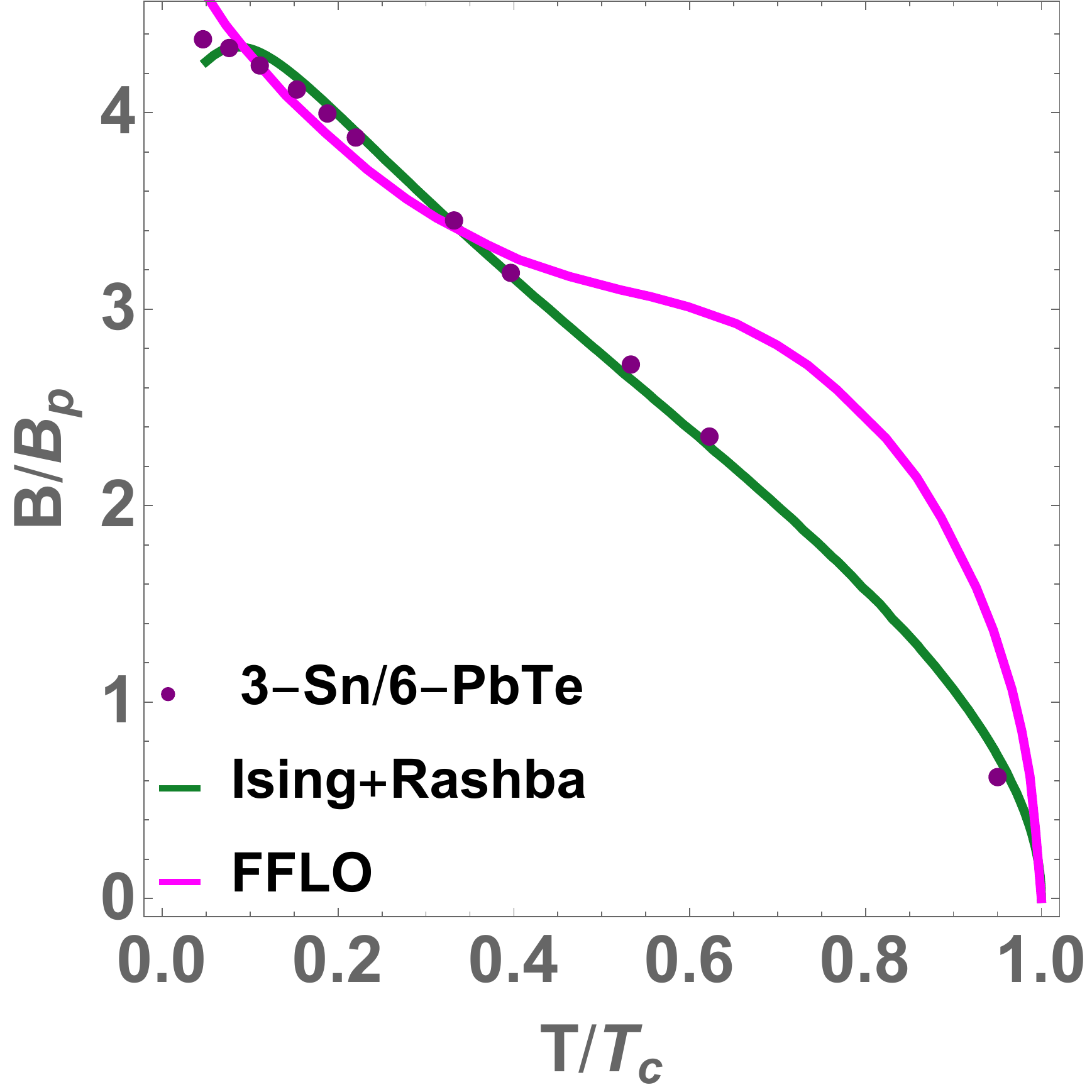}
		\caption{(Color online) The best fitting curve in the low temperature regime with the formula 2D FFLO state\cite{PhysRevB.50.12760} is represented by purple line. The fitting curve of Ising pairing is shown by the green line. The dots represent the $B_C$ data of few-layer stanene 3-Sn/6-PbTe, $T_{c}=0.45 \mathrm{K}$ and $B_{p}=0.84 \mathrm{T}$, and the experimental data is from from Ref.\cite{falson2019type}.
		The fitting curves show that the 2D FFLO phase is not quantitatively consistent with the data\cite{falson2019type}. %Remark to Hongchao: the fitting and data need to be changed to 3-Sn/6-PbTe for higher B/Bp and recheck the parameters after the revision. 
		}\label{fitting}
	\end{figure}

\bibliography{Ising}